# DARCS: Memory-Efficient Deep Compressed Sensing Reconstruction for Acceleration of 3D Whole-Heart Coronary MR Angiography

Zhihao Xue, Fan Yang, Juan Gao, Zhuo Chen, Hao Peng, Chao Zou, Hang Jin, and Chenxi Hu, *Member, IEEE*

*Abstract*—Three-dimensional coronary magnetic resonance angiography (CMRA) demands reconstruction algorithms that can significantly suppress the artifacts from a heavily undersampled acquisition. While unrolling-based deep reconstruction methods have achieved state-of-the-art performance on 2D image reconstruction, their application to 3D reconstruction is hindered by the large amount of memory needed to train an unrolled network. In this study, we propose a memory-efficient deep compressed sensing method by employing a sparsifying transform based on a pre-trained artifact estimation network. The motivation is that the artifact image estimated by a well-trained network is sparse when the input image is artifact-free, and less sparse when the input image is artifact-affected. Thus, the artifact-estimation network can be used as an inherent sparsifying transform. The proposed method, named De-Aliasing Regularization based Compressed Sensing (DARCS), was compared with a traditional compressed sensing method, de-aliasing generative adversarial network (DAGAN), model-based deep learning (MoDL), and plug-and-play for accelerations of 3D CMRA. The results demonstrate that the proposed method improved the reconstruction quality relative to the compared methods by a large margin. Furthermore, the proposed method well generalized for different undersampling rates and noise levels. The memory usage of the proposed method was only 63% of that needed by MoDL. In conclusion, the proposed method achieves improved reconstruction quality for 3D CMRA with reduced memory burden.

## I. INTRODUCTION

THREE-DIMENSIONAL coronary magnetic resonance angiography (3D CMRA) is an imaging technique that has demonstrated significant potential for the assessment and diagnosis of coronary artery disease (CAD). This non-invasive method provides clear visualization of coronary arteries without subjecting patients to ionizing radiation [1], making it valuable for early detection and long-term monitoring of stenosis in the coronary arteries [2, 3]. However, one significant drawback of 3D CMRA is its relatively long scan time, typically ranging from 7 to 10 minutes, which is significantly longer than 2D cardiac MR imaging [4-6]. This poses a substantial obstacle to its widespread clinical application, prompting the need for reconstruction algorithms capable of substantially accelerating 3D CMRA acquisition.

To date, two types of methods have been developed to address this challenge: classic Bayesian reconstruction approaches and deep learning-based techniques. Classic Bayesian methods leverage the "sparsity" [7, 8] or "low-rankness" [9] of the image to improve conditioning of the otherwise ill-posed reconstruction problem. Among these methods, PROST utilizes self-consistency between image patches at different locations, and has achieved state-of-the-art performance for undersampled CMRA reconstruction [5, 10]. However, due to the non-specificity of the compressed sensing (CS) priors, images reconstructed using these methods often suffer from residual aliasing artifacts, especially at high acceleration rates.

Recently, deep learning approaches have emerged as a potent strategy for undersampled image reconstruction. Early studies focused on developing networks that directly map undersampled data to reconstructed images, with notable methods including AUTOMAP [11], U-Net [12, 13] and generative adversarial network (GAN) based methods [14-16] such as De-Aliasing GAN (DAGAN) [14]. While these methods have shown promise, their pure data-driven nature lacks interpretability, and their generalizability remains challenging due to limited training data in the field of medical image reconstruction, especially in CMRA. Therefore, recent works have aimed to combine the strength of model-driven data-fidelity constraints with data-driven deep learning-based regularization into a single approach. This has given rise to two

This work was partially supported by the National Natural Science Foundation of China (No. 62001288) and the Shanghai Science and Technology Commission (No. 22TS1400200) (Corresponding author: Chenxi Hu).

Zhihao Xue, Fan Yang, Juan Gao, Zhuo Chen, and Chenxi Hu are with the National Engineering Research Center of Advanced Magnetic Resonance Technologies for Diagnosis and Therapy, School of Biomedical Engineering, Shanghai Jiao Tong University, Shanghai, China (e-mail: thomasxue@sjtu.edu.cn; fanyang035@sjtu.edu.cn; gaojuan_student@sjtu.edu.cn; chenzhuo_sjtu@sjtu.edu.cn; chenxi.hu@sjtu.edu.cn).

Hao Peng and Chao Zou are with Shenzhen Institutes of Advanced Technology, Chinese Academy of Sciences, Guangdong, China (e-mail: hao.peng@siat.ac.cn; chao.zou@siat.ac.cn).

Hang Jin is with Department of Radiology, Zhongshan Hospital, Fudan University and Shanghai Medical Imaging Institute, Shanghai, China (email: jin.hang@zs-hospital.sh.cn)

well-known strategies: Plug-and-play (PnP) and unrolling. PnP [17, 18], initially developed without incorporating deep learning-based priors, involves replacing the proximal operator in each iteration with either an off-the-shelf denoiser or a pre-trained denoising network, resulting in enhanced reconstruction compared to standard CS methods [18, 19]. However, since the denoising regularizer is not specific to the imaging artifacts, PnP can still suffer from residual artifacts at a high acceleration rate. Deep unrolling utilizes an end-to-end trained unrolled network with two components used recursively, where the first maintains data fidelity and the second enforces regularization based on a convolutional neural network (CNN). Several representative unrolled networks including ADMM-Net [20], ISTA-Net [21], MoDL [22], and E2E-VarNet [23] have achieved the state-of-the-art performance in 2D MRI reconstruction. A few of them have been applied to 3D CMRA [4, 6]. However, the use of unrolling for 3D image reconstruction has been limited by the large memory burden in the training process [6], leading to suboptimal reconstruction quality.

To address these limitations, several new deep learning reconstruction strategies have been developed. Deep equilibrium (DEQ) model [24-27] extends unrolling by running the iteration to convergence, so that the backward propagation can be performed with respect to a single layer, which substantially reduced the memory burden. However, to evaluate the gradient in the back propagation, the iterations need to reach the fixed point, which could substantially increase the computational time. Another recently proposed method [28] combines a deep de-aliasing network and an iterative CS algorithm. The method generates an artifact-reduced image based on a pre-trained de-aliasing network, and then forces the final reconstruction to stay close to the artifact-reduced image while satisfying the data-fidelity constraint. Since the network is pre-trained, the memory burden of the method is also reduced.

In this study, we aim to propose a novel deep compressed sensing method that has both a low memory burden and a task-specific regularization, so that it can accurately reconstruct the image for highly accelerated 3D CMRA. Similar to the method in [28], a pre-trained de-aliasing network is used to generate the image prior. However, the de-aliasing network is used to provide a sparsifying transform, rather than a static image for the subsequent compressed sensing iterations. The motivation is that the artifact image estimated by a de-aliasing network is sparse when the input image is artifact-free, and less sparse when the input image is artifact-affected. Thus, the de-aliasing network inherently induces a sparsifying transform for the underlying task. The proposed method, named De-Aliasing Regularization based Compressed Sensing (DARCS), was compared with PROST, DAGAN, MoDL, and plug-and-play for accelerations of 3D CMRA in a cohort of 10 healthy subjects.

## II. RELATED WORKS

### A. Classical Reconstruction Methods for CMRA

Classical reconstruction methods, including parallel imaging, compressed sensing, and low-rank based methods, have been the mainstream for acceleration of CMRA. Piccini *et al* proposed a CMRA technique based on a 3D radial trajectory and XD-GRASP [29] reconstruction, which employs sparsity of the images along time respiratory dimension [30]. This method has also been extended to free-running 5D imaging [31], which reconstructs all cardiac and respiratory phases. Bustin *et al* proposed a patch-based low-rank method, called 3D-PROST, which showed improved image quality than standard wavelet-based compressed sensing method [5, 32]. Before the advent of deep learning, PROST was widely considered the state of the art in this field. The limitation for these classic acceleration methods is that they use hand-crafted regularizations, which are empirically assumed for medical images, such as smoothness and low-rankness. However, medical images are more than just being smooth and low-ranked. They also form a unique distribution different from natural images. Furthermore, the distribution of image artifacts is also important since learning of it may more precisely help the reconstruction to remove these artifacts. Therefore, the failure to learn application-specific prior about the images and artifacts make these classical methods less effective in the presence of high accelerations.

### B. Data-driven Deep Learning Methods

Early deep learning reconstruction methods are purely data-driven. A network is often constructed to directly transform undersampled k-space or an aliased image into an aliasing-free image. For example, Zhu *et al* constructed a neural network to transform k-space data undersampled by different patterns into an aliasing-free image [11]. Hyun *et al* developed a U-Net method to reconstruct the aliasing-free image from keyhole-undersampled k-space [12]. Yang *et al* proposed a conditional generative adversarial network for MRI reconstruction [14], which combined the adversarial loss with so-called content loss, which includes pixelwise $l2$ loss and the perceptual loss [33]. Data-driven deep learning methods have also been used in CMRA, but mainly for denoising [34] and super-resolution [35]. A main limitation for data-driven approaches is the ignorance of k-space data and the imaging model. At the testing phase, even though the reconstructed images may not satisfy the data-fidelity constraint, this information is ignored and not utilized to improve the reconstruction quality [22].

### C. Model-driven Deep Learning Methods

The model-driven deep learning methods were proposed to mitigate the above issues associated with pure data-driven methods. Model-driven deep learning methods can be broadly classified into three categories. The first category is the so-called plug-and-play approaches. PnP was initially proposed without the use of deep learning [17]. In the traditional PnP, an off-the-shelf denoiser replaces the proximal step in a conventional alternating direction algorithm, which has been empirically observed to generate better reconstruction results. The exact reason for this improvement, however, is not clear and still subject to debate [18]. Zhang *et al* firstly combined PnP with deep learning [36], by using a denoising CNN in the replacement. This strategy was then extended by several other groups for different reconstruction tasks [37, 38]. The second category is the unrolling method. Unrolling abandoned the use of pre-trained networks; instead, it unrolls a reconstruction algorithm into a deep network, which often comprises alternations of the data-fidelity module and a CNN [22]. The network is then trained end-to-end to transform an aliasing-affected image into an aliasing-free image, much like the data-driven approaches, which learned the task-specific information related to the image and artifacts. Furthermore, since data-fidelity is recursively enforced, unrolling exhibits a higher degree of generalizability than pure data-driven approaches. The third category is methods based on generative priors. In these methods, a prior is trained with a variety of approaches, such as variational autoencoder [39, 40], PixelCNN++ [41], score-based diffusion [42-44], and energy-based modeling [45]. It was often found that these methods have better generalizability than discriminative methods, which were trained from pairs of aliasing-affected/aliasing-free images, on out-of-distribution data [42, 44]. However, for in-distribution data, the performance of these methods appeared to be similar or inferior to unrolling [44, 45]. Furthermore, the reconstruction or training of these methods was typically quite time-consuming or memory-consuming [43], rendering them difficult for 3D reconstructions. So far, only unrolling has been used in CMRA.

Although deep unrolling has shown superior performance compared to classical compressed sensing, a major limitation for unrolling in 3D imaging applications is the excessive usage of the GPU memory during the training phase [26, 46, 47]. Since an algorithm is unrolled and trained end-to-end, the intermediate variable in each layer need to be stored in memory. For high-resolution 3D imaging, the memory usage becomes formidable, which limits the maximal number of unrolled iterations typically to 3 or 4. This truncation of unrolled iterations compromises the reconstruction quality [26]. To address this issue, deep equilibrium model based methods [24-27, 48] have been proposed. These methods can perform potentially an infinite number of iterations in forward pass by calculating the fixed point. Moreover, in the backward pass, the gradient can be obtained without backpropagating through the large number of forward iterations, eliminating the need to store the intermediate variables. As a result, these methods alleviate the memory burden during training by only requiring the memory necessary to compute the fixed point and gradient. However, the iterative solving procedure is seriously time-consuming, and in some studies [27] the DEQ method even occupied more memory for solving the inverse of the operator in backpropagation. Another type of approach, proposed by Liao *et al*. [28], is an AI-assisted compressed sensing which combines deep learning and iterative reconstruction. In this framework, an image inferred by a deep de-aliasing network is used as a spatial regularizer to guide iterative CS reconstruction, without the need for large amounts of memory during training. However, the use of the $l$1 norm between the reconstruction and a fixed de-aliased image, which may not be accurate, may introduce bias into the final reconstruction due to the oversimplified prior.

Finally, although interpretability of unrolling is better than pure data-driven approaches, it is difficult to assess *ad hoc* how the reconstruction would be before the input was sent to the network. It is also unclear what kind of image prior is used for an unrolled network [45]. These limitations motivated us to develop a compressed sensing method with a prior obtained by a separate discriminative training instead of an unrolled end-to-end training. We show that this method not only preserves the task-specificity of the image prior, but also reduces memory usage and improves interpretability.

### III. METHODOLOGY

#### A. Problem Formulation

In MRI reconstruction, the undersampled k-space measurement $y \in \mathbb{C}^m$ is associated with the underlying image $x \in \mathbb{C}^n$ through the following equation:

$$y = Ax + \epsilon \quad (1)$$

where the operator $A \in \mathbb{C}^{m \times n}$ is the forward model of the imaging system and $\epsilon$ the noise. In the presence of multiple coils, the forward model is often a composition of three linear operators:

$$A = DFS \quad (2)$$

where $S$ is the sensitivity map of coils, $F$ the Fourier transform, and $D$ the undersampling operator in k-space. Due to ill-conditioning of the inverse problem, the standard CS reconstruction often involves formulating and solving the following optimization problem:

$$\underset{x}{\arg\min} \|y - Ax\|_2^2 + \alpha \|\Phi(x)\|_1 \quad (3)$$

where $\|y - Ax\|_2^2$ is the data fidelity term, $\Phi$ the sparsifying transform, and $\alpha$ the regularization weight. Traditionally, the regularization term is usually based on transform-domain sparsity [7, 49, 50] or low-rankness [9, 10, 51], which are hand-crafted priors for the image. For highly undersampled k-space, these traditional compressed sensing methods often lack regularization power, resulting in aliasing artifacts or blurred reconstructions [14].

#### B. De-aliasing Regularization Based Compressed Sensing

To mitigate the above issues, we propose to learn a task-specific regularizer based on deep convolutional neural networks, which does not need unrolling of the algorithm and an end-to-end training of the unrolled network. Fig. 1 shows a schematic of the proposed method. To use the method, we firstly train a de-aliasing network, denoted by $G(x; \widehat{\theta})$, where $\widehat{\theta}$ represents the learned parameters of the network. The model $G(x; \widehat{\theta})$ thus aims to remove the aliasing artifacts in $x$, and the residual $G(x; \widehat{\theta}) - x$ aims to estimate the artifact of the image. If the image is aliasing-free, $G(x; \widehat{\theta}) - x$ should output a zero image, which intrinsically has the highest sparsity. If the image is aliasing-affected, $G(x; \widehat{\theta}) - x$ should output an image of recognized artifacts, which cause a reduction of the sparsity.

$$x^* = \arg\min_{x} \|y - Ax\|_2^2 + \alpha \|G(x; \hat{\theta}) - x\|_1$$

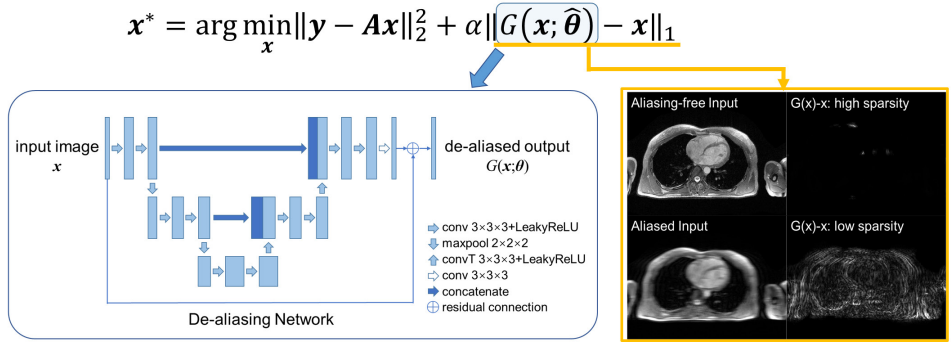

Fig. 1 An overview of the De-Aliasing Regularization based Compressed Sensing (DARCS) method. In this framework, the generator in a pre-trained 3D de-aliasing generative adversarial network (architecture shown in the figure) is used to build the regularizer for compressed sensing. The rationale is that the residual output G(x)-x has a high sparsity for an aliasing-free image and a low sparsity for an aliasing-affected image, thereby providing an intrinsic sparsifying transform for this underlying image. When the residual connection is used in the de-aliasing network as shown here, the output before the residual connection is automatically the residual.

Therefore, the mapping $x \to G(x; \hat{\theta}) - x$ serves as an intrinsic sparsifying transform, and its $l1$-norm can be used as a regularizer for the underlying image reconstruction. That is, once the network is trained, we can minimize the following cost function to estimate the underlying $x$:

$$\arg\min_{x} \|y - Ax\|_2^2 + \alpha \|G(x; \hat{\theta}) - x\|_1 \quad (4)$$

An advantage for the de-aliasing-based regularization relative to traditional regularizations is the improvement of task specificity. To see it, we show the sparsity map of a trained DARCS regularizer, the finite difference regularizer, and the wavelet regularizer for both the zero-filled images and ground truth in Fig. 2. Whereas the DARCS regularizer has a low sparsity for the zero-filled images and high sparsity for the ground truth, the other two traditional regularizers have closer degrees of sparsity for the two types of images, since edges are unspecifically penalized by these regularizers.

### C. Optimization

The optimization problem in Eq 4 is solved by the Alternating Direction Method of Multipliers (ADMM), which iteratively executes the following three steps:

Step 1:
$$x_t = \arg\min_{x} \|y - Ax\|_2^2 + \mu \|x - z_{t-1} + u_{t-1}\|_2^2 \quad (5a)$$

Step 2:
$$z_t = \arg\min_{z} \alpha \|G(z; \hat{\theta}) - z\|_1 + \mu \|x_t - z + u_{t-1}\|_2^2 \quad (5b)$$

Step 3:
$$u_t = u_{t-1} + (x_t - z_t) \quad (5c)$$

where $u_t$ is the scaled Lagrange multiplier, $z_t$ the variable introduced in variable-splitting, and $\mu$ the penalty weight. Among the three steps, Step 1 is a regularized SENSE reconstruction problem, which can be solved by a conjugate-gradient (CG) algorithm; Step 2 is the proximal step of the ADMM algorithm; Step 3 is an update of the Lagrangian multiplier. We solve Step 2 by a gradient-descent (GD) algorithm [41], where the gradient at each iteration is given by

$$g_k = 2\mu(z - (x_t + u_{t-1})) + \alpha(J - I)^T \text{sign}(G(z; \hat{\theta}) - z) \quad (6)$$

where $J$ denotes Jacobian of $G$ with respect to $z$, and $I$ denotes an identity matrix. Algorithm 1 shows a pseudo code for the entire ADMM algorithm.

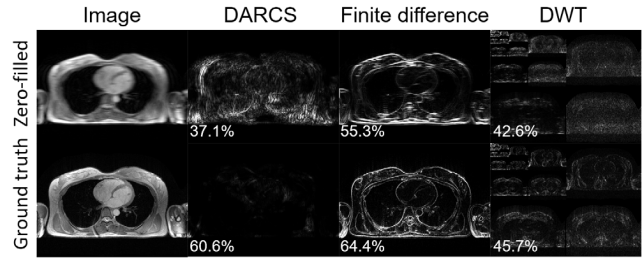

Fig. 2 The outputs of each sparsifying transform for the zero-filled and ground truth images. Compared to finite difference and discrete wavelet transform (DWT), the artifact-estimation-based sparsifying transform generates more distinguishable degrees of sparsity between the zero-filled and ground truth images. The number in each image denotes the percentage of pixels with the intensity less than 10% of the maximum, which roughly quantifies sparsity of the image.

---

**Algorithm 1** De-aliaisng regularization based on compressed sensing (DARCS).

**Input:** $x_0 = A^H y, z_{0,0} = 0, u_0 = 0, k = 0$
**Output:** $z_T$
1: **for** $t = 1 : T$ **do**
2: $\quad x_t = (A^H A + \mu I)^{-1}[A^H y + \mu(z_{t-1,k} - u_{t-1})]$
$\quad$ // solve using conjugate gradient method
3: $\quad z_{t,0} = x_t + u_{t-1}$
4: $\quad$ **for** $k = 1 : K$ **do**
5: $\quad\quad g_{t,k} = \frac{d}{dz}[\mu \|z - z_{t,0}\|_2^2 + \alpha \|G(z; \hat{\theta}) - z\|_1]\big|_{z=z_{t,k-1}}$
6: $\quad\quad z_{t,k} = z_{t,k-1} - \beta g_{t,k}$ // gradient descent
7: $\quad$ **end for**
8: $\quad u_t = u_{t-1} + x_t - z_t$
9: **end for**

---

### D. Choice of the De-aliasing Network

Theoretically, any network trained to de-aliase the images can be used to construct the proposed regularization. However, different network architectures, loss functions, or training strategies may generate different de-aliasing performance, which in turn causes different regularization performances. For example, it is known that the use of adversarial loss often improves the perceptual quality of the network outputs in image-to-image translation tasks [52]. Hence, in this work, we trained a DAGAN [14] as the de-aliasing network. The architecture of the DAGAN generator is shown in Fig. 1. The generator was based on a 3D U-Net with a residual connection between the output and the input. Two input and output channels were used to store the real and imaginary parts of the input and output, respectively. The discriminator of the DAGAN was based on a CNN, which comprised of 5 convolutional blocks and 1 fully-connected layer, followed by a sigmoid function. The network was trained with the same loss function as in [14], which comprised of adversarial loss, the image-domain normalized mean-squared-error (NMSE) loss, the frequency-domain NMSE loss (the frequency loss), and the VGG perceptual loss [33].

### E. Training Strategies

Normally, a de-aliasing network is trained using pairs of aliasing-affected images and the corresponding aliasing-free images. We refer to this routine training strategy as Strategy 1. The use of this strategy, however, does not guarantee the network would output an artifact-free image when the input is also artifact-free. Since we need not only low sparsity for aliasing-affected images, but also high sparsity for aliasing-free images, we developed Strategy 2 (shown in supplementary Figure S1 (http://arxiv.org/abs/2402.00320), which uses both aliasing-affected images and aliasing-free images to train the network. During training, both types of images were input with a probability of 0.5. Finally, both Strategy 1 and 2 only trained a single de-aliasing network. However, during the ADMM iterations, the intermediate images may have somewhat different distributions than both the initial aliasing-affected images and the aliasing-free images. Therefore, for Strategy 3, we trained two DAGANs, each processing the aliased images generated at different phases of the ADMM algorithm. More specifically, we trained the first DAGAN by the same data as that used in Strategy 2. Then, we used the induced regularizer to reconstruct images from the training dataset using the ADMM algorithm with 10 iterations. We then formed a new set of aliased images based on these intermediate results, and trained the second DAGAN with these images and the aliasing-free images with an equal probability. Compared with Strategy 2, each DAGAN in Strategy 3 can be more specific to the aliasing artifacts at the corresponding phase during the iterations.

## IV. EXPERIMENTS AND RESULTS

### A. CMRA Data Collection and Preparation

The study was approved by the Institutional Review Board (IRB). All subjects provided written informed consent prior to the scan. Twenty healthy subjects were scanned on a 3.0T MRI scanner (uMR 790, United Imaging Healthcare Co. Ltd., Shanghai, China) with a 12-channel torso coil and 32-channel spine coil. Three-dimensional CMRA was acquired with an electrocardiogram-triggered, navigator-gated, T2-prepared dual-echo spoiled gradient echo sequence [53] with the following parameters: repetition time (TR) = 5.21ms, echo time (TE) = 2.24/3.17ms, FOV = 400mm×300mm×90mm, flip angle = 10°, resolution = 1.10mm×1.10mm×1.50mm, undersampling factor=2, and bandwidth =1070 Hz/pixel. The central 24×24 lines were densely sampled. Prior to reconstruction, all k-space data were compressed to 12 virtual coils [54] using Berkeley Advanced Reconstruction Toolbox(BART) [55]. Coil sensitivity maps were estimated using ESPIRiT [56] from the central 24×24 lines. The ground truth images were then reconstructed with iterative SENSE. These ground truth images were normalized to have a maximal magnitude of 1. Since there were two echoes, 40 volumes were reconstructed from the 20 subjects, among whom 10 randomly chosen subjects were used for training while the remaining 10 were used for testing. These ground truth images were used to synthesize undersampled k-space based on a pseudorandom Poisson-disc trajectory [57] at an acceleration factor of 8 and the coil sensitivity maps. The iterative SENSE reconstruction of these undersampled data was then used as the input to the DAGAN.

### B. Implementation Details

Both the network and reconstruction were implemented with PyTorch (version 1.11.0). Training was performed with ADAM [58] with an initial learning rate of $1\times10^{-4}$, which was reduced by half every 32,000 iterations. To obtain large number of training samples and fit the GPU memory, the images were cropped along the fully sampled readout dimension with size of 20. Extra data augmentation was not performed, following conventions from several previous studies [4, 35]. The adversarial loss, NMSE loss, frequency loss, and VGG perceptual loss were applied with a weighting of 1, 15, 0.1, and 0.025, respectively. When the training Strategy 1 and Strategy 2 were used, the ADMM reconstruction algorithm was run for 20 iterations, with $\alpha=0.1$, $\mu=0.005$, $K=2$, and step size $\beta=0.01$ in Algorithm 1. When the training Strategy 3 was used, the ADMM algorithm was run for 20 iterations, with the first 10 iterations using the parameters above, and the second 10 iterations using $\alpha=0.1$, $\mu=0.01$, $K=2$, and $\beta=0.01$. All training was performed on a GPU workstation equipped with a NVIDIA RTX 3090 GPU with a 24GB memory.

### C. Evaluation of Regularization Performance

We performed experiments to evaluate the proposed regularization method when different de-aliasing models, training strategies, and regularization parameter values were used. Results of these experiments show the relationship between these parameters and the performance of the resultant regularization.

#### 1) The Impact of De-aliasing Models

We employed both a DAGAN and a regular 3D U-Net trained with only the NMSE loss for construction of the regularizer. The two networks had the same architecture and training except the change of the loss function. Supplementary Figure S2 shows the iterative NMSE reductions associated with the two regularizers during the reconstruction process. DAGAN exhibited slightly lower NMSE compared with U-Net (0.0140 vs. 0.0144, paired $t$-test P<0.01). Hence, we used DAGAN as the de-aliasing model in further experiments.

#### 2) The Impact of Training Strategies

We employed the three training strategies for training of the de-aliasing model. Fig. 3 shows the iterative NMSE reductions associated with the three strategies. It is clear that Strategy 3 led to the lowest NMSE at the final iteration due to the use of stage-adaptive regularizers. Strategy 2 led to slightly lower NMSE than Strategy 1 due to inclusion of the aliasing-free images in the training dataset, which generated more contrastive sparsities between the initializer and the ground truth image. All following experiments used Strategy 3 for training the de-aliasing models.

### 3) The Impact of Regularization Parameters

We tested the reconstruction algorithm with a variety of regularization weights to see if the algorithm is robust against such changes. The results are shown in Fig. 4. The changes of NMSE from $\alpha = 0.005$ to $\alpha = 1$ were small, and the minimal NMSE was achieved at $\alpha = 0.1$. The results suggest that the reconstruction algorithm had a stable performance across mild variations of the regularization weight.

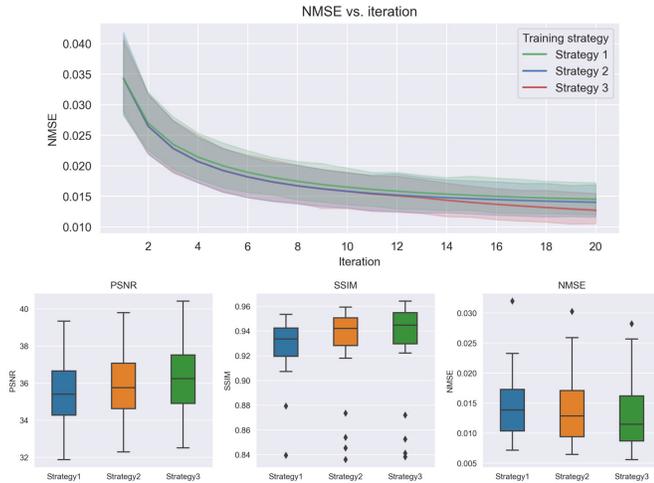

Fig. 3 The impact of different training strategies. The iterative NMSE reductions associated with the three strategies were compared. PSNR, SSIM, and NMSE of final reconstructions were also compared for each sample. Results show that Strategy 3 led to the lowest NMSE among the three strategies (paired *t*-test, all P < 0.05).

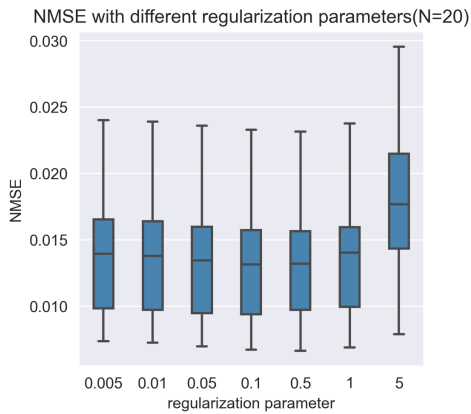

Fig. 4 NMSE of DARCS reconstruction with different regularization parameters. The change of NMSE with respect to $\alpha$ indicates the insensitivity to the selection of regularization parameter in a certain range.

### D. Comparison to State-of-the-arts

In this section, we compared our method to a variety of reconstruction methods, including PROST, PnP, 3D MoDL, and DAGAN. PROST was implemented using the open-source code in Ref [9]. The following parameters were used after tuning: patch size = 5, search window = 20, number of similar patches selected = 20, patch offset = 4, and regularization parameter = $5\times10^{-3}$. The PnP method was implemented with the 3D denoising CNN (DnCNN) [59] as suggested by Ref [18]. The 3D DnCNN was trained to recover the original ground truth images from those with added Gaussian noise (standard deviation < 0.2). The MoDL was modified to support 3D convolution and pooling layers. Due to the limited GPU memory, only 3 unrolled iterations were used in the network, which is consistent with a recent work using unrolling for CMRA reconstruction[6]. Each CNN module consisted of five 3D convolution blocks with shared weights, as suggested by the original model[22]. The DAGAN used the same architecture and loss function as those used by the proposed method.

### 1) In-distribution Comparisons

We firstly compared these methods with in-distribution testing data, which were prepared with the same k-space undersampling mask at an 8-fold acceleration rate. Fig. 5 shows representative reconstructions from the five studied methods. Both the coronal and transversal views of the reconstructed images and their errors with respect to the ground truth are displayed. The reconstructed images from PROST had residual blurring and artifacts compared with the ground truth, and had the lowest PSNR, SSIM and the highest NMSE among all 5 methods. PnP, MoDL, and DAGAN improved PSNR, SSIM, and reduced NMSE relative to PROST. However, there were visible remaining artifacts in the results of PnP and MoDL, and the results of DAGAN appeared slightly over-smoothed. The proposed DARCS method exhibited the highest image qualities in both coronal and transversal views and the best PSNR, SSIM and NMSE among all 5 methods.

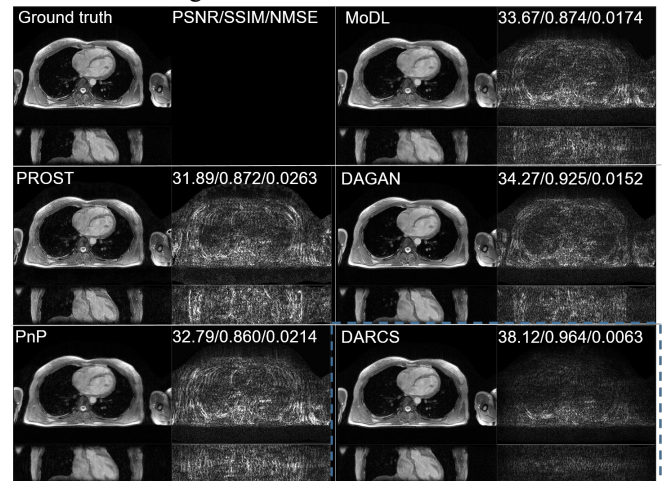

Fig 5. Comparisons of reconstructions from PROST, PnP, 3D MoDL, DAGAN, and DARCS in both coronal and transversal views of one subject. The acceleration rate was eight. The error maps were also presented. DARCS improved the aliasing artifact suppression relative to the other methods.

TABLE I
QUANTITATIVE COMPARISONS BETWEEN DIFFERENT METHODS IN TERMS OF PSNR, SSIM, AND NMSE VALUES AT 8-FOLD ACCELERATION.

| Method | PSNR(dB) | SSIM | NMSE |
| --- | --- | --- | --- |
| Zero-filled | 27.16±1.74 | 0.690±0.049 | 0.0970±0.0424 |
| PROST | 30.43±1.87 | 0.836±0.040 | 0.0459±0.0228 |
| PnP | 31.34±1.67 | 0.820±0.033 | 0.0365±0.0146 |
| MoDL | 31.97±1.76 | 0.849±0.034 | 0.0321±0.0141 |
| DAGAN | 32.73±1.94 | 0.873±0.049 | 0.0273±0.0134 |
| DARCS | **36.13±2.01** | **0.928±0.042** | **0.0127±0.0058** |

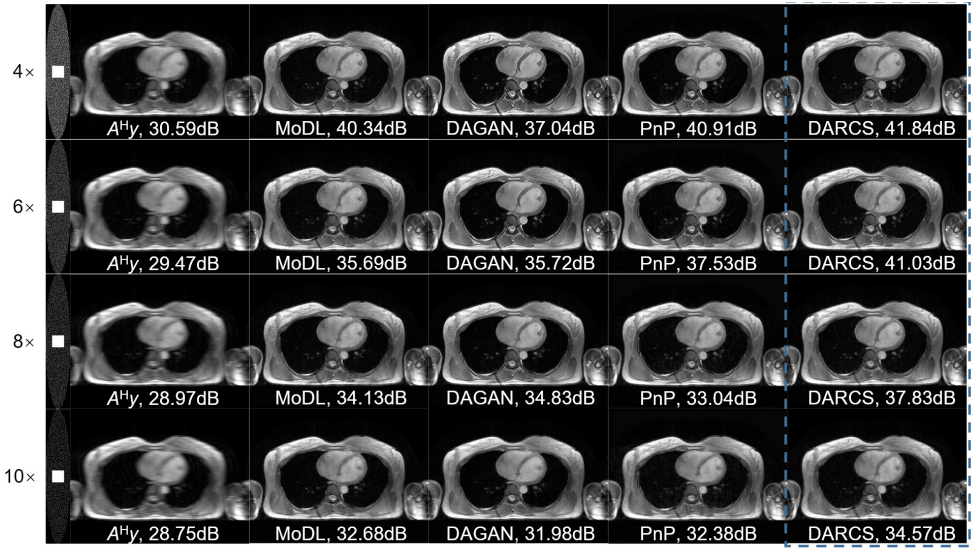

Fig.6 Reconstructions of DARCS, MoDL, PnP, and DAGAN at different undersampling rates of 4, 6, 8, and 10. Note that all these methods were trained on the 8-fold undersampled data. MoDL, PnP, and DAGAN showed oversmoothing and increased artifacts at higher acceleration rates, which were less noticeable in the DARCS reconstructions. DARCS consistently provided a reconstruction of a higher quality than the other methods across all acceleration rates.

Table I shows a statistical comparison of these methods over 10 healthy subjects. While all 5 methods led to significantly improved statistics than the zero-filled reconstructions, DARCS significantly improved PSNR, SSIM, and reduced NMSE over the other methods by a large margin. These results suggest that DARCS led to superior reconstruction accuracies compared with these methods.

### 2) Out-of-distribution Comparisons

The undersampling rate of the acquisition may change since the need for acceleration may vary depending on the body size, imaging resolution, and imaging time of the underlying scan. Thus, we compared the studied methods with variable undersampling rates between 4 and 10. Notice that all algorithms were trained at the 8-fold acceleration. Thus, data from other acceleration rates are considered out-of-distribution, because the severity of the aliasing artifacts can be different.

Fig. 6 shows the reconstruction results from all 5 studied methods for a representative subject. The k-space undersampling mask was shown on the left. All methods led to higher accuracies when the acceleration rate was reduced and lower accuracies when the acceleration rate was increased. However, DARCS outperformed other methods at every acceleration rates. Table II shows a statistical comparison of the four deep learning methods from R=4 to R=10. For all the acceleration rates, DARCS showed statistically better PSNR, SSIM, and NMSE over the other methods. These results suggest that DARCS has a reasonable generalizability across different acceleration rates.

### 3) Robustness to noise

To investigate the robustness of DARCS to noise, we conducted experiments using measurements with added white Gaussian noise in k-space. The noise level ($l$) was determined by the percentage of the norm of the noise to the norm of the k-space data, i.e. $l = \|\epsilon\|_2/\|y\|_2$. We trained a DAGAN model at R=8 with 2% noise. Fig. 7 compares the reconstructions of SENSE, DAGAN and the corresponding DARCS on two subjects of the test dataset when the data contains 2% and 5% noise. Both the de-aliasing network and our method showed tolerance to noise, considering the quantitative metrics and image quality. However, DARCS consistently outperformed DAGAN, demonstrating its superior performance.

### 4) Interpretability of DARCS

The interpretability of DARCS is improved compared with DAGAN, PnP, or MoDL because sparsity of each iterate can be visualized. This visualization not only confirms whether sparsity has a consistent improvement, but also informs us where the artifacts roughly reside in the image space. An example is shown in the supplementary Figure S3, where the sparsity maps at iterations 1, 11, and 20 are shown for regularizers obtained from the 3 training strategies. For unrolling, DAGAN, and PnP, this information is not accessible, since they do not have an explicit definition of the image prior.

TABLE II
QUANTITATIVE COMPARISONS OF PSNR/SSIM/NMSE FOR 3D MoDL, DAGAN, PnP, AND DARCS AT DIFFERENT UNDERSAMPLING RATES (R).

| Method | R | PSNR(dB) | SSIM | NMSE |
|---|---|---|---|---|
| MoDL | 4 | 39.06±2.85 | 0.899±0.029 | 0.0068±0.0039 |
|  | 6 | 33.92±1.96 | 0.872±0.036 | 0.0206±0.0095 |
|  | 8 | 31.97±1.76 | 0.849±0.034 | 0.0321±0.0141 |
|  | 10 | 30.46±1.72 | 0.807±0.034 | 0.0455±0.0202 |
| DAGAN | 4 | 33.15±1.60 | 0.933±0.029 | 0.0240±0.0091 |
|  | 6 | 32.95±1.77 | 0.900±0.036 | 0.0254±0.0108 |
|  | 8 | 32.73±1.94 | 0.873±0.049 | 0.0273±0.0134 |
|  | 10 | 31.09±1.93 | 0.836±0.042 | 0.0398±0.0196 |
| PnP | 4 | 38.58±1.63 | 0.908±0.034 | 0.0071±0.0034 |
|  | 6 | 35.48±1.65 | 0.868±0.034 | 0.0141±0.0057 |
|  | 8 | 31.34±1.67 | 0.820±0.033 | 0.0365±0.0146 |
|  | 10 | 30.70±1.58 | 0.777±0.035 | 0.0418±0.0148 |
| DARCS | 4 | **43.33±3.17** | **0.965±0.033** | **0.0029±0.0026** |
|  | 6 | **41.12±2.78** | **0.963±0.030** | **0.0043±0.0028** |
|  | 8 | **36.13±2.01** | **0.928±0.042** | **0.0127±0.0058** |
|  | 10 | **32.37±1.71** | **0.877±0.033** | **0.0290±0.0122** |

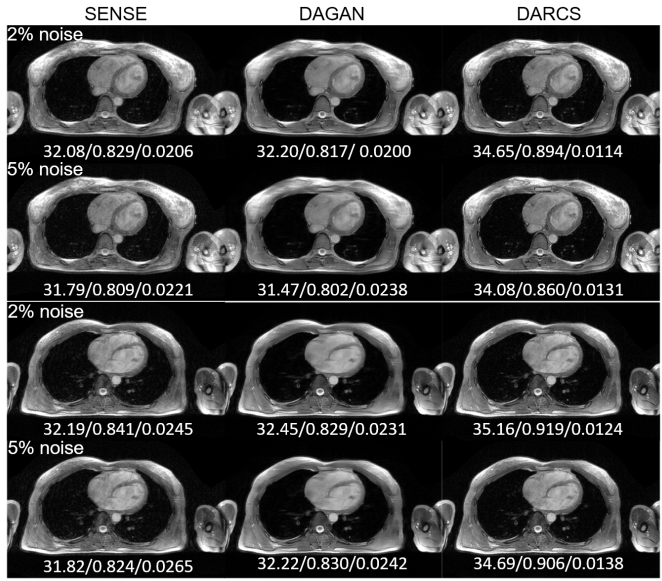

Fig.7 Reconstruction results of SENSE, DAGAN and DARCS from noisy measurements with different noise levels (2% and 5%). The values in the bottom are the PSNR/SSIM/NMSE values of each volume.

### 5) Comparisons of Technical Metrics

Another advantage of the proposed method relative to unrolling is that it reduces the memory burden associated with training. Reduction of the memory burden also allows the network to have more parameters and become deeper, which may further improve the de-aliasing performance. Table III shows the comparison of technical metrics, including number of parameters, memory usage during training, training time, and testing time, of different methods. DARCS has more parameters but lower memory consumption than MoDL. The training is slightly longer for DARCS, because the network is more complex. The testing time is much longer for DARCS compared with MoDL or DAGAN, which is a limitation of the iterative technique. However, the testing time is still much shorter than traditional compressed sensing method, such as PROST.

TABLE III
COMPUTING TIME WHEN APPLYING EACH METHOD TO ONE VOLUNTEER (TWO-ECHO IMAGE) IN THE TEST DATASET (10 PEOPLE IN TOTAL). BATCH SIZES WERE UNIFIED TO COMPARE MEMORY USAGE.

| Method | Number of parameters | Memory usage | Training time | Testing time |
| --- | --- | --- | --- | --- |
| PROST | -- | -- | -- | 45.1±6.9 min |
| MoDL | 0.34M | 22.9 GB | 27.2 h | 33.1±1.5 s |
| PnP | 0.05M | 10.8 GB | 14.6 h | 212.5±7.2s |
| DAGAN | 1.55M(G) 7.36M(D) | 14.4 GB | 39.3 h | 1.4±0.3 s |
| DARCS | 1.55M(G) 7.36M(D) | 14.4 GB | 39.8 h | 255.2±1.5s |

### E. Visualization of the Coronary Arteries

We reformatted the images to generate a visualization of the right coronary artery (RCA) and left anterior descending (LAD) artery. A two-point Dixon water-fat separation method [60] was applied to the dual-echo 3D images to suppress signals from the epicardial fat. To visualize the coronary artery, maximum intensity projection (MIP) was performed with a slice thickness of 6 mm using 3D Slicer (Version 5.2.2) [61]. The results for one representative subject are shown in Fig. 8(a), where DARCS showed the best reconstruction of the two arteries among all 5 studied methods. All the other methods obscured the distant segment of the two arteries due to the undersampling (indicated by orange arrows). In Fig. 8(b), we analyzed the maximum visible vessel lengths of 10 testing subjects. The vessel lengths were not significantly different between the ground truth and DARCS reconstructions (paired $t$-test, RCA: P=0.33>0.05; LAD: P=0.55>0.05). These findings indicate that DARCS reconstructions yield a satisfactory image quality for visualization of the coronary arteries.

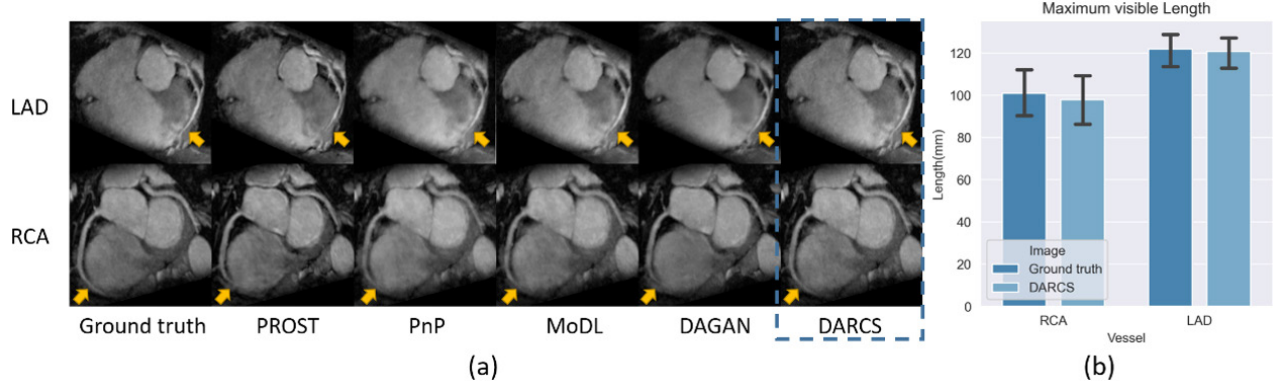

Fig.8 (a) Visualization of the right coronary artery and left anterior descending artery using the MIP reformation. DARCS generated a clear and sharp visualization of the coronary artery, which was consistent with the ground truth. On the other hand, blurring and loss of the distal vessel segments can be observed for the other methods (arrows in the figure). (b) Lengths of RCA and LAD in 10 test subjects were statistically compared between the ground truth and DARCS reconstructions using paired $t$-test. The differences were not significant (RCA: P=0.33; LAD: P=0.55)

## V. DISCUSSION

This work presents a compressed sensing method for 3D CMRA reconstruction based on a learned sparsifying transform, which is the difference between the output of a 3D de-aliasing network and its input. The learned sparsifying transform has a task-specific regularization capability and a low memory burden during training. The method demonstrated superior performance on reducing aliasing artifacts and restoring fine structures at large acceleration rates, outperforming other methods such as PnP, 3D MoDL, DAGAN, and PROST. In addition, experiments indicate that a stage-adaptive training further improves reconstruction quality. Although only two stages were employed, adding more stages into the stage-adaptive training is feasible and may further improve the performance. Finally, coronary artery images obtained by using DARCS at an 8-fold acceleration still maintained high quality, suggesting the potential for clinical translation.

Due to the pre-training with paired artifact-affected and artifact-free images, the proposed image prior can precisely penalize the underlying artifacts, resulting in a superior performance relative to the hand-crafted regularization and plug-and-play reconstruction. The incorporation of the data-fidelity constraint renders the method also model-based, which improves its generalizability for unseen data compared with purely data-driven methods, such as DAGAN. Note that even though the same DAGAN was used to construct the image prior in DARCS, the latter outperformed the former, suggesting the inclusion of the image model is valuable. Finally, compared to unrolling, the method has a substantially lower memory burden, making it more feasible for 3D high-resolution reconstruction tasks. The reduced memory burden facilitates the use of a more complex network, which may eventually translate to a better reconstruction, as found in this work.

Compared with recent methods aiming to reduce the memory burden associated with unrolling [24-27, 48], our method also has several important differences. For example, our method does not require training a network with respect to the fixed point, which is required for DEQ. The latter causes a significantly increase of the training time and hinders its application to 3D CMRA. In [28], the authors also trained a de-aliasing network and incorporated it into the CS framework. However, whereas the method in [28] only employs the de-aliasing network to generate a fixed aliasing-reduced image for the subsequent CS reconstruction, our method uses the de-aliasing network itself to provide a sparsifying transform. Thus, in the referenced method, the proxy provided by the de-aliasing network remains unchanged during iteration, whereas in our method, the proxy would dynamically change over successive iterations. Since the initial guess of the image from the de-aliasing network based on the zero-filled image may not be accurate, dynamically evolving the solutions based on a combination of the data-fidelity constraint and the learned sparsifying transform may facilitate a more accurate reconstruction.

Our approach and study have some limitations. Like other iterative algorithms, our method also requires a relatively long computational time for reconstruction. However, the current reconstruction time is within a reasonable range for the 3D CMRA application and substantially faster than traditional CS methods like PROST. Secondly, due to the difficulty in collecting a large number of 3D CMRA samples, our study is limited to a relatively small dataset. Future studies based on a larger multi-center cohort and prospective undersampling are needed to fully verify the performance of the proposed method.

## VI. Conclusion

In this paper, we propose a novel compressed sensing method based on a learned sparsifying transform, which is generated from a pre-trained de-aliasing network. The method exhibits improved image quality over state-of-the-art methods, a lower memory burden than deep unrolling, and a faithful depiction of the coronary arteries despite an 8-fold acceleration. The method can be used to either reduce the acquisition time or increase the spatial resolution for 3D CMRA.

# Supporting Information

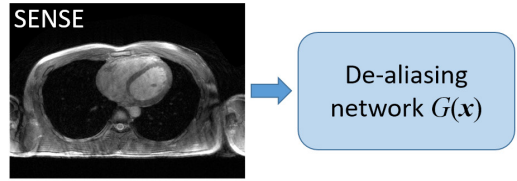
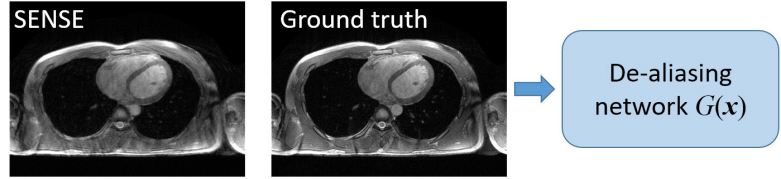
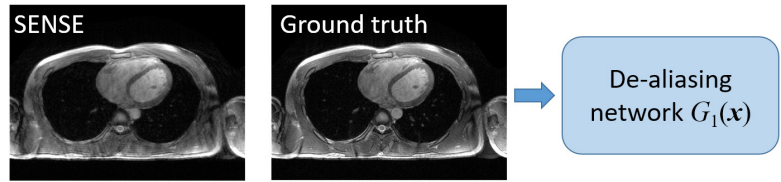
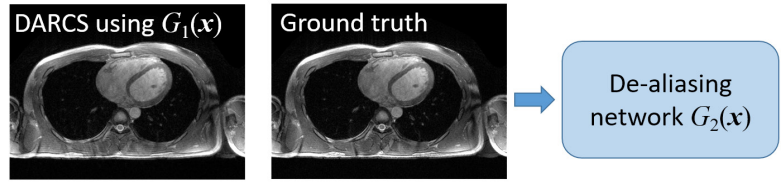

Figure S1. Training strategies for obtaining the de-aliasing regularizer. In Strategy 1, SENSE reconstructions, which still contain some undersampling artifacts, are used to pre-train the de-aliasing network, while in Strategy 2 the ground truth images are added to the input images. In Strategy 3, the reconstructions obtained by DARCS with network in Strategy 2 and ground truth images are used to train another de-aliasing network. The target outputs of all the three strategies are the ground truth images.

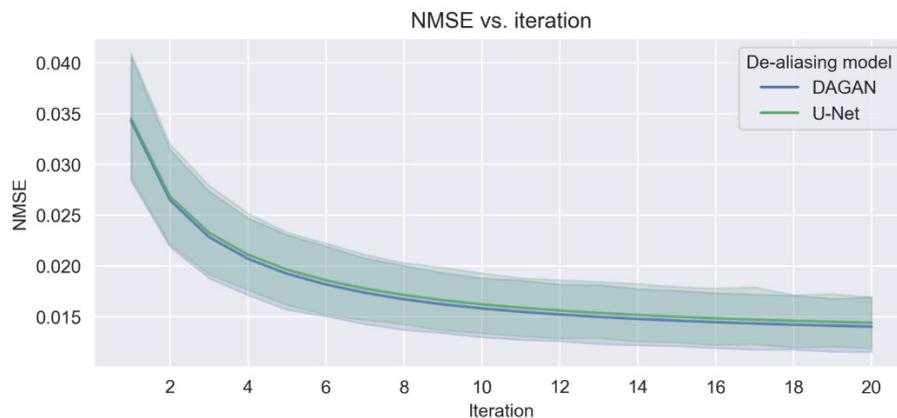

Figure S2 The comparison of NMSE reductions during iteration using different de-aliasing models (DAGAN or U-Net) as the regularizer in DARCS. A slight difference between the results of the two models can be observed, with DAGAN exhibiting lower NMSE.

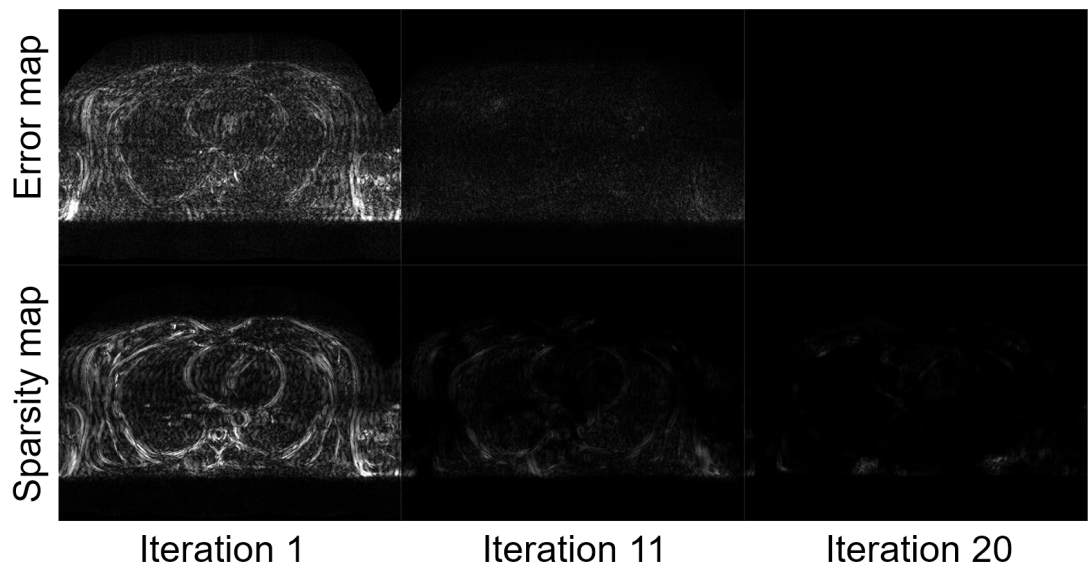

Figure S3. The error map and sparsity map at iterations 1, 11, and 20. The error map was generated by subtracting the iterate from the ground truth. The sparsity map was the output of the learned sparsifying transform, which informs us about the error recognized by the sparsifying transform.